\documentclass[aps,prb,10pt,twocolumn,amsmath,amssymb,longbibliography,superscriptaddress,nofootinbib]{revtex4-1}

\usepackage{amsmath,amssymb,amsfonts} 	
\usepackage{graphicx}
\usepackage{hhline}
\usepackage[latin1]{inputenc}
\usepackage{subfigure}

\begin{document}

\title{Calculating critical temperatures for ferromagnetic order in two-dimensional materials.}

\author{Daniele Torelli}
\affiliation{CAMD, Department of Physics, Technical University of Denmark, 2820 Kgs. Lyngby Denmark}
\author{Thomas Olsen}
\email{tolsen@fysik.dtu.dk}
\affiliation{CAMD, Department of Physics, Technical University of Denmark, 2820 Kgs. Lyngby Denmark}

\begin{abstract}
Magnetic order in two-dimensional (2D) materials is intimately coupled to magnetic anisotropy (MA) since the Mermin-Wagner theorem implies that rotational symmetry cannot be spontaneously broken at finite temperatures in 2D. Large MA thus comprises a key ingredient in the search for magnetic 2D materials that retains the magnetic order above room temperature. Magnetic interactions are typically modeled in terms of Heisenberg models and the temperature dependence on magnetic properties can be obtained with the Random Phase Approximation (RPA), which treats magnon interactions at the mean-field level. In the present work we show that large MA gives rise to strong magnon-magnon interactions that leads to a drastic failure of the RPA. We then demonstrate that classical Monte Carlo (MC) simulations correctly describe the critical temperatures in the large MA limit and agree with RPA when the MA becomes small. A fit of the MC results leads to a simple expression for the critical temperatures as a function of MA and exchange coupling constants, which significantly simplifies the theoretical search for new 2D magnetic materials with high critical temperatures. The expression is tested on a monolayer of CrI$_3$, which were recently observed to exhibit ferromagnetic order below 45 K and we find excellent agreement with the experimental value.
\end{abstract}
\maketitle

\section{Introduction}
In 2017 it was demonstrated that a monolayer of CrI$_3$ exhibits ferromagnetic order with a Curie temperature of 45 K.\cite{Huang2017a} The discovery comprises the first example of magnetism in a two-dimensional (2D) material and it was subsequently shown that the magnetic properties of few layers of CrI$_3$ can be controlled with an electric field\cite{Jiang2018, Huang2018, Huang2018a, Jiang2018a} and act as spin-filters.\cite{Song2018, Klein2018a} The observation  has initiated a vast amount of interest in the subject\cite{Sachs2013a, Zhuang2016a, McGuire2017, Seyler2017, Lado2017, Miao2018, Ersan2018} and a few other 2D materials have been shown to exhibit promising magnetic properties as well. For example, it has been shown that ferromagnetic order persists down to the bilayer limit in Cr$_2$Ge$_2$Te$_6$\cite{Gong2017b} and room temperature magnetic order has been observed for monolayers of VSe$_2$ on a van der Waals substrate.\cite{Bonilla2018} However, the description of magnetic order in 2D is more demanding than in three dimensions and it is currently a difficult task to predict the critical temperature of a given material. While it is well-known that magnetic order in 2D is driven by magnetic anisotropy (MA) there is no simple descriptor for the critical temperature and one has to rely on complex simulations in order to obtain reliable estimates. 

Magnetic order in crystalline solids is an inherently correlated effect that arises as a consequence of Pauli exclusion and electron-electron interactions. The theoretical description thus comprises a highly challenging topic and a series of approximations are required in order to derive various quantities related to the magnetic properties of solids. Typically, the problem is mapped onto a Heisenberg model of the form\cite{Yosida1996}
\begin{equation}\label{eq:heisenberg}
 H=-\frac{1}{2}\sum_{ij}J_{ij}\mathbf{S}_i\cdot\mathbf{S}_j,
\end{equation}
where $\mathbf{S}_i$ is the spin operator at site $j$ and $J_{ij}$ are the magnetic coupling between spins at site $i$ and $j$, that account for both direct exchange and superexchange.\cite{Anderson1950,Anderson1959} Still, a quantum mechanical treatment of Eq. \eqref{eq:heisenberg} is non-trivial and the eigenstates cannot be obtained by analytical means in general. A direct numerical treatment is also out of the question due to the vast Hilbert space required for solids.

If one is interested in the critical temperature, a simple expression can be obtained from mean field theory, where it is assumed that each spin only couples to the average magnetic field of the crystal. In the ferromagnetic case where all sites carry the same spin one obtains\cite{Yosida1996}
\begin{equation}\label{eq:T_mf}
 T^{\mathrm{MF}}_c=\frac{S(S+1)}{3k_B}J_0,
\end{equation}
where $k_B$ is Boltzmanns constant, $S$ is the maximum eigenvalue of $S_z$ and $J_0=\sum_jJ_{ij}$. Correlation effects are completely neglected in mean field theory and the present expression does not in general yield quantitative agreement with experiments. However, for three-dimensional materials it does provide a rough estimate of $T_c$ and has been widely applied to predict critical temperature and to extract exchange coupling constants from measured values of $T_c$.\cite{Samuelsen1971,Ami1995,Bose2010,Johnston2011}  In contrast, the Mermin-Wagner theorem implies that magnetic order in a 2D material cannot persist at finite temperatures unless MA is present. Since the derivation of Eq. \eqref{eq:T_mf} does not make any assumptions of dimensionality of the problem, a direct consequence of the Mermin-Wagner theorem is that mean field theory cannot be applied in 2D. The main purpose of the present work is to obtain a simple equivalent of Eq. \eqref{eq:T_mf} that can be applied to obtain the Curie temperature of 2D ferromagnetic materials for a given set of anisotropy parameters.

Another approach to analyzing the Heisenberg model is based on a Holstein-Primakoff transformation of the spin operators. In that case the Hamiltonian can be written as power series in bosonic field operators and the leading quadratic part is then straightforward to diagonalize. The excitations can be interpreted as non-interacting spin-waves and provide a good approximation for the spectrum at low temperatures. The remaining terms in the Hamiltonian represent spin-wave interactions and need to be taken into account at finite temperatures where many spin-waves are typically present. This is not a trivial task, however, but one can include the quartic term in a Hartree-Fock type of approximation, whihch gives rise to a temperature dependent renormalization of the spin-waves. In the present work we will refer to this as the Random Phase Approximation (RPA),\cite{Yosida1996} but we note that this terminology is also sometimes used for the time-dependent susceptibility in the Hartree-Fock approximation. For 3D materials Curie temperatures obtained from RPA provide a more accurate estimate than those obtained from mean field theory and tend to provide a lower bound for the exact value, whereas mean field theory provides an upper bound.\cite{Rusz2006, Bergqvist2004} More importantly, RPA respects the Mermin-Wagner theorem and gives rise to vanishing critical temperatures in the absence of magnetic anisotropy. RPA thus appears to comprise the simplest method that can provide quantitative agreement with experiment. However, as will be shown in the present work, the RPA fails dramatically in systems with large anisotropy. For 2D materials it is expected that large anisotropy is exactly the property needed if one is searching for materials with high critical temperatures and RPA is not a suitable approximation in such cases.

A third approach to the problem is based on the fact that at large temperatures quantum effects tends to be quenched by thermal fluctuations and one can consider a classical approximation for the model. Critical temperatures can then be obtained by performing Monte Carlo simulations of the model at different temperatures and identify the point where the average magnetization vanishes. While such an approach includes all correlation in the model, it is in general difficult to assess the importance of the neglected quantum effects. In particular for systems with $S=1/2$ quantum effects are likely to be important even at elevated temperatures. Nevertheless, classical Monte Carlo simulations have been shown to provide excellent agreement with experimental critical temperatures for diluted magnetic semiconductors \cite{Bergqvist2004} and Heusler alloys \cite{Rusz2006}. Moreover, the classical treatment has the important property that it correctly approaches the Ising limit for large anisotropies. 

The classical simulations are, however, rather demanding in terms of computational load and it would be highly desirable to have an analytical expression that replaces Eq. \eqref{eq:T_mf} for two-dimensional materials with anisotropy. In the present paper we obtain such an expression by fitting the results of Monte Carlo simulations in anisotropic two-dimensional systems to an analytical expression. We perform the simulations for honeycomb, quadratic, and hexagonal lattices and provide a universal expression that only depends on the number of nearest neighbors and the critical temperature of the corresponding Ising model. We then consider a monolayer of CrI$_3$ as a test example and obtain good agreement with the experimental value of $T_c$ using Heisenberg parameters obtained from first principles calculations.

\section{Theory}\label{sec:theory}
The starting point of our calculations is the Heisenberg model with nearest neighbor exchange interactions $J$, single ion anisotropy $A$, and nearest neighbor anisotropic exchange $B$. Typically, 2D materials are isotropic in-plane and magnetic order is therefore only possible if the easy axis (here chosen as the $z$-axis) is perpendicular to the plane of the material. We thus consider the model Hamiltonian
\begin{align}\label{eq:H}
H=-\frac{1}{2}\sum_{ij}J_{ij}\mathbf{S}_i\cdot\mathbf{S}_j-A\sum_i(S^z_i)^2-\frac{1}{2}\sum_{ij}B_{ij}S^z_iS^z_j,
\end{align}
with $J_{ij}$, $A, B_{ij}>0$. The sums run over all magnetic sites and $J_{ij}=J$, $B_{ij}=B$ if $i$ and $j$ are nearest neighbors and zero otherwise. The maximum value of $S_i^z$ is denoted by $S$. Without MA, all eigenenergies are proportional to $J$ and the model does not contain any fundamental interaction parameters. This implies that the amount of correlation only depends on the lattice and the value of $S$. In contrast, when MA is present, the last term in Eq. \eqref{eq:H} introduces additional correlations, which is quantified by the dimensionless coupling constants $A/J$ and $B/J$. Importantly, it should be noted that in the limit of $A/J\rightarrow\infty$, all excitations will have spins aligned along the easy axis and the model becomes equivalent to the Ising model with coupling parameter $J^{\mathrm{Ising}}=J+B$.

The critical temperature is defined as the temperature at which the magnetic order vanishes. For a ferromagnetic system it can be determined by calculating the magnetization as
\begin{align}
M(T)=\frac{1}{Z}\sum_{\{s\}}M_se^{-E_s/k_BT},
\end{align}
where $Z$ is the partition function, $s$ denotes eigenstates of Eq. \eqref{eq:H}, and $M_s$ are the corresponding magnetic moments. From Eq. \eqref{eq:H} it is clear $T_c/J$ must be a function of $A/J$, $B/J$, and $S$ as well as the lattice. A common approach to obtain approximate solutions to Eq. \eqref{eq:H} is the Holstein-Primakoff transformation that replaces the spin operators by bosonic raising and lowering operators.\cite{Bruno1991, Yosida1996} The Hamiltonian can then be written as
\begin{align}\label{eq:H_ex}
H=E_0 + S\tilde H_0+\tilde H_1+\frac{1}{S}\tilde H_2+\ldots
\end{align}
where $E_0$ is the ground state energy. $S\tilde H_0$ is quadratic in raising and lowering operators, $\tilde H_1$ is quartic in raising and lowering operators, $\tilde H_2$ contains sixth order terms and so forth. The terms beyond $\tilde H_0$ thus introduce interactions between the Holstein-Primakoff bosons. We note that the anisotropy constants $A$ and $B$ only enters in $E_0$, $\tilde H_0$ and $\tilde H_1$.  With periodic boundary conditions, all excitations can be labeled by a Bloch momentum $\mathbf{q}$ and all terms in the Hamiltonian can be written in terms of $a^\dag_{\nu\mathbf{q}}$ and $a_{\nu\mathbf{q}}$, which create and annihilate Holstein-Primakoff bosons for the sublattice $\nu$ at wavevector $\mathbf{q}$. Since the bosons carry spin-1, the magnetization per site can be written as
\begin{align}\label{eq:M}
M(T)=M_0 - \frac{1}{nN_\mathbf{q}}\sum_{n\mathbf{q}}n_B(E_{n\mathbf{q}},T),
\end{align}
where $M_0$ is the ground state magnetization per site, $N_\mathbf{q}$ is the number of unit cells, $n$ is a band index, $n_B$ is the Bose distribution, and $E_{n\mathbf{q}}$ are the eigenenergies of the Hamiltonian.

For simplicity we will restrict ourselves to a single site per unit cell in the following. If one neglects the interactions, the dispersion is readily obtained yielding
\begin{align}\label{eq:eps0}
\varepsilon_{\mathbf{q}}=\varepsilon_{\mathbf{q}}^0+A(2S-1)+SBN_{nn},
\end{align}
where $N_{nn}$ is the number of nearest neighbors and $\varepsilon_{\mathbf{q}}^0$ is the dispersion of spinwaves without anisotropy, which satisfies $\varepsilon_{\mathbf{0}}^0=0$. Magnetic order at finite temperature in 2D is only possible if the spectrum is gapped and thus depends on the presence of MA. The expression becomes exact in the limit of vanishing temperature and we see that the single-ion anisotropy will introduce a gap in the spectrum if $S>1/2$. For $S=1/2$ the single-ion anisotropy alone does not introduce a gap and finite critical temperatures are only possible with non-vanishing anisotropic exchange. In general magnetic order will be possible if $A(2S-1)+SBN_{nn}>0$. The interacting part of the anisotropy terms in the Hamiltonian \eqref{eq:H_ex} to fourth order in the field operators becomes
\begin{align}\label{eq:H_int}
\tilde H_1^{\mathrm{Ani}}=-\frac{1}{2N_\mathbf{q}}\sum_{\mathbf{q}\mathbf{q}'\mathbf{q}''}[2A+\tilde B(\mathbf{q}''-\mathbf{q})]a^\dag_\mathbf{q'}a^\dag_\mathbf{q''}a_\mathbf{q}a_\mathbf{q+q'-q''}
\end{align}
with
\begin{align}
\tilde B(\mathbf{q})=B\sum_{j}\cos(\mathbf{q}\cdot\mathbf{R}_j),
\end{align}
where the sum runs over the set of smallest lattice vectors. The single-ion anisotropy thus introduces attractive interactions between the Holstein-Primakoff bosons, whereas the sign of the anisotropic exchange interaction depends on the value of $\mathbf{q}$.

Whereas $\tilde H_0$ is readily diagonalized, the interaction terms require some level of approximation. Taking the Hartree-Fock approximation for $\tilde H_1$ leads to the Random Phase Approximation (RPA), which has previously been shown to provide good estimates of the Curie temperatures in 3D where the MA is usually neglected. The temperature dependent corrections to the spectrum without anisotropy $\Delta\varepsilon_\mathbf{q}$ is well-known and can be found in Ref. \onlinecite{Yosida1996}. In the presence of anisotropy the RPA spectrum (for a single site per unit cell) acquires the additional temperature dependent terms
\begin{align}\label{eq:H_rpa}
\Delta\varepsilon_\mathbf{q}^{\mathrm{Ani}}&=-\frac{1}{N_\mathbf{q}}\sum_{\mathbf{q}'}\Big[4A+\tilde B(\mathbf{q}-\mathbf{q}')+\tilde B(\mathbf{0})\Big]n_B(E_{\mathbf{q}'}^{\mathrm{RPA}},T)
\end{align}
with $E_{\mathbf{q}}^{\mathrm{RPA}}=\varepsilon_{\mathbf{q}}+\Delta\varepsilon_\mathbf{q}+\Delta\varepsilon_\mathbf{q}^{\mathrm{Ani}}$. The RPA dispersion and resulting magnetization at a given temperature thus have to be calculated self-consistently. It should be mentioned that the procedure of calculating the critical temperature as the point where Eq. \eqref{eq:M} vanishes appears to be ill-defined in the RPA, since the renormalized spin gap approached zero before the magnetization vanishes. In particular, if we take $B=0$ it can be seen from Eqs. \eqref{eq:eps0} and \eqref{eq:H_rpa} that the gap should close when $\langle n\rangle=(2S-1)/4$, which is always smaller than $M_0=S$. However, the average number of bosons in the system diverges as the gap approaches zero, and the magnetization is not well-defined at this point. Nevertheless, since the average number of bosons diverge as the gap closes we have that $dM(T)/dT\rightarrow-\infty$ as $\Delta\varepsilon_\mathbf{0}+\Delta\varepsilon_\mathbf{0}^{\mathrm{Ani}}\rightarrow0$. We can thus calculate the magnetization up to arbitrarily small values of the spin gap, where the magnetization acquires a large negative slope as a function of temperature and we may simply identify the critical temperature as the point where the renormalized spin wave gap closes.

In the case of large MA the RPA provides qualitatively wrong results since the Hartree-Fock level of theory cannot correctly capture the strong correlation introduced by the additional interaction term. This is inherited from the non-interacting spin-wave theory, where the breakdown can be seen as follows: assuming a quadratic dispersion of the form $\varepsilon_q=aq^2+\Delta$ the magnetization becomes 
\begin{align}
M(T)=&M_0-\frac{L^2}{(2\pi)^2}\int\frac{d^2q}{e^{(aq^2+\Delta)/k_BT}-1}\notag\\
=&M_0-k_BT\frac{L^2}{\pi a}\int_{\Delta/k_bT}\frac{dx}{e^{x}-1}\notag\\
=&M_0+k_BT\frac{L^2}{\pi a}\log(1-e^{-\Delta/k_BT}).
\end{align}
At the critical temperature the magnetization vanishes, which implies
\begin{align}
1-e^{-\Delta/k_bT_c}=e^{-\alpha/k_BT_c},
\end{align}
where $\alpha=\pi aM_0/L^2$. This equation allows for arbitrarily large solutions for $T_c$ if $\Delta$ is taken large enough (high anisotropy). In RPA one obtains a similar picture, since the average treatment of interactions simply introduces a temperature dependent rescaling of $\Delta$. The fact that $T_c$ diverges for large values of the anisotropy signals a breakdown of the RPA, since for a fixed value of $B$ the critical temperature has to approach the Ising limit asymptotically as we take $A\rightarrow\infty$.

\section{Results}
\subsection{Single-ion anisotropy}
\begin{figure}
	\includegraphics[width=0.5\textwidth]{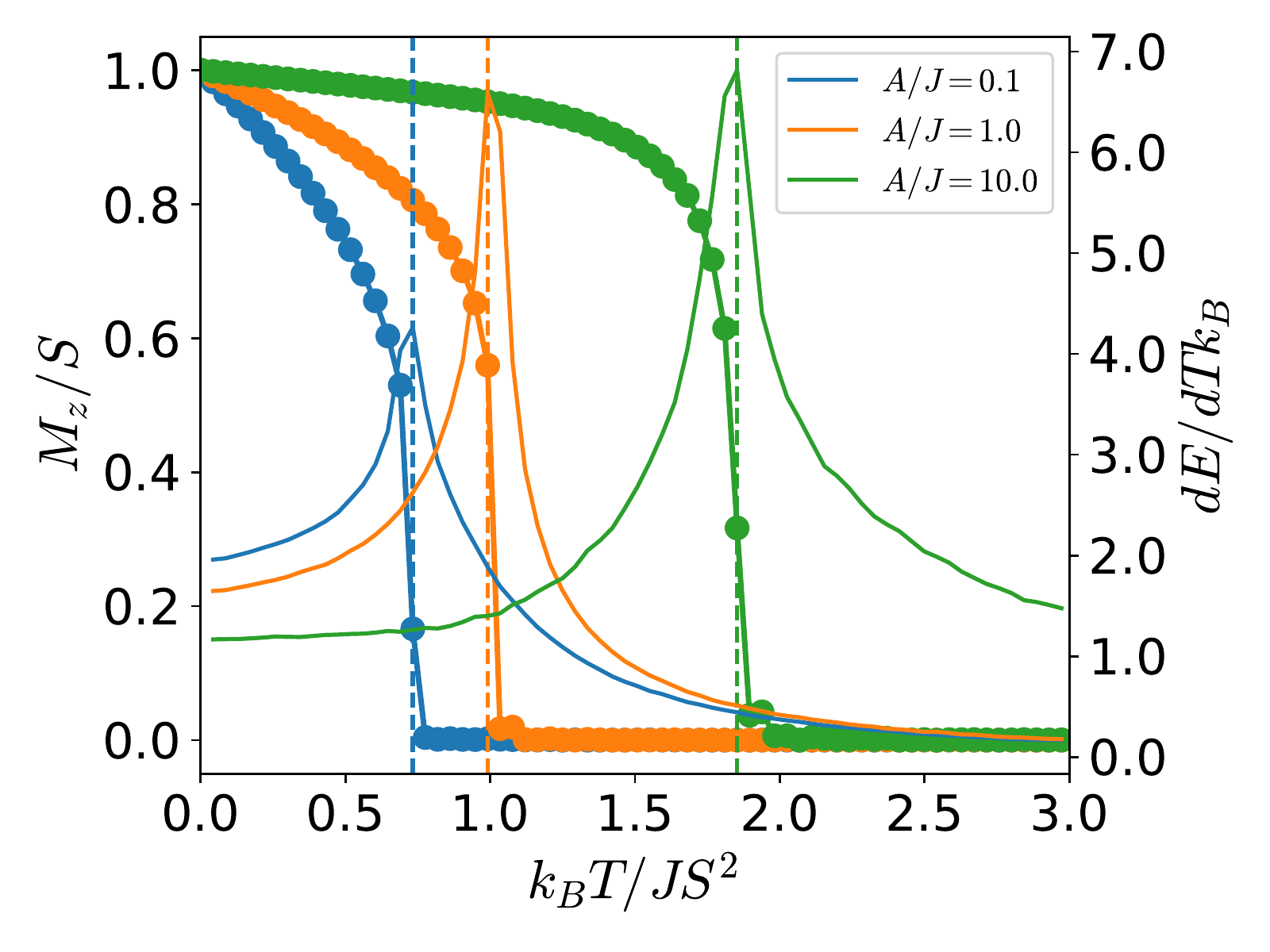}
	\caption{Magnetization and heat capacity $C=dE/dT$ as a function of temperature for the square lattice with three different values of $A/J$. The critical temperatures are indicate by dashed vertical lines.}
	\label{fig:magnetization}
\end{figure}

We now turn to the Monte Carlo simulations of the classical Heisenberg model at different temperatures. We start by considering only single-ion anisotropy and thus set $B=0$. It is expected that the classical treatment is unreliable at low temperature where quantum fluctuations may dominate but at large temperatures the quantum fluctuation will be quenched by thermal fluctuations. A major advantage of this approach is the fact that the Ising limit is naturally satisfied and the classical approximation will thus become asymptotically exact when $A/J\rightarrow\infty$ and $S\neq1/2$. In Fig. \ref{fig:magnetization} we show examples of the MC simulations for the square lattice using three different values of $A/J$. The critical temperature can be extracted as  the point where the magnetization vanishes or as the temperature where the heat capacity $C=dE/dT$ diverges. The two-approaches gives identical results for the critical temperature.

\begin{figure}
	\includegraphics[width=0.5\textwidth]{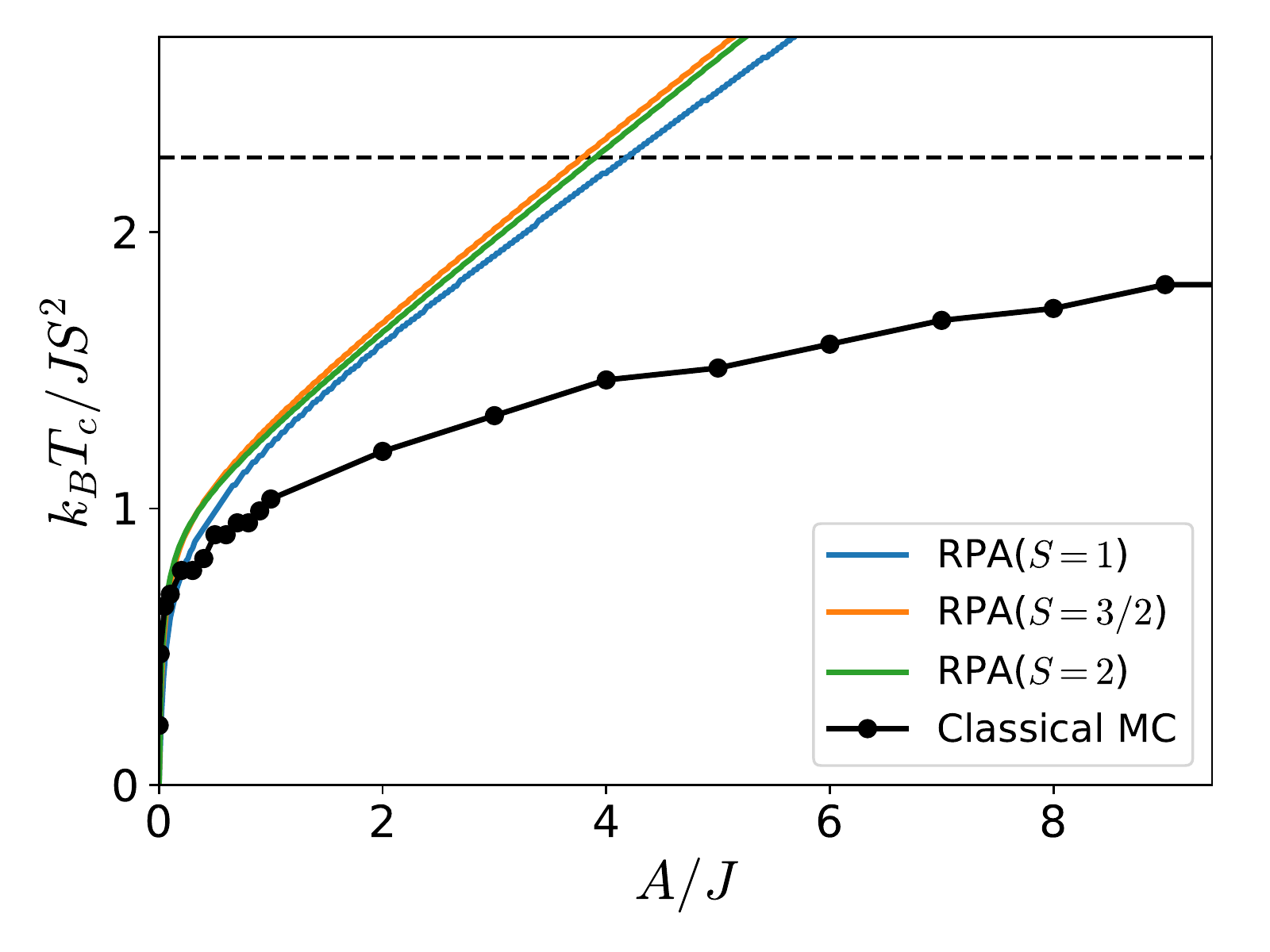}
	\caption{Critical temperature for a quadratic lattice as a function of rescaled single-ion anisotropy $A/J$ calculated for $S=1$, $S=3/2$, and $S=2$ with ferromagnetic exchange coupling using the RPA. The Ising critical temperature comprises an exact upper limit and is indicated by dashed line.}
	\label{fig:rpa}
\end{figure}

In Fig. \ref{fig:rpa} we compare the critical temperature obtained from the RPA with classical MC simulations on a quadratic lattice. For the classical model the critical temperature can be written as $k_BT_c^{\mathrm{Cl}}=S^2Jf_{\mathrm{Cl}}(A/J)$, where $f$ is a universal function that do not depend on $S$. For the RPA one has $k_BT_c^{\mathrm{RPA}}=S^2Jf_{\mathrm{RPA}}(S, A/J)$, but it is clear from Fig. \ref{fig:rpa} that $f_{\mathrm{RPA}}(S, A/J)$ is nearly independent of $S$. The RPA clearly violates the Ising limit as noted above and we expect that $T_c^{\mathrm{RPA}}$ becomes unreliable when $A\sim J$. The Ising limit is approached rather slowly as $A/J$ is increased and we can conclude that the materials with large anisotropies that will typically be referred to as Ising type ferromagnets are poorly described by the Ising model. 

\begin{figure}
	\includegraphics[width=0.5\textwidth]{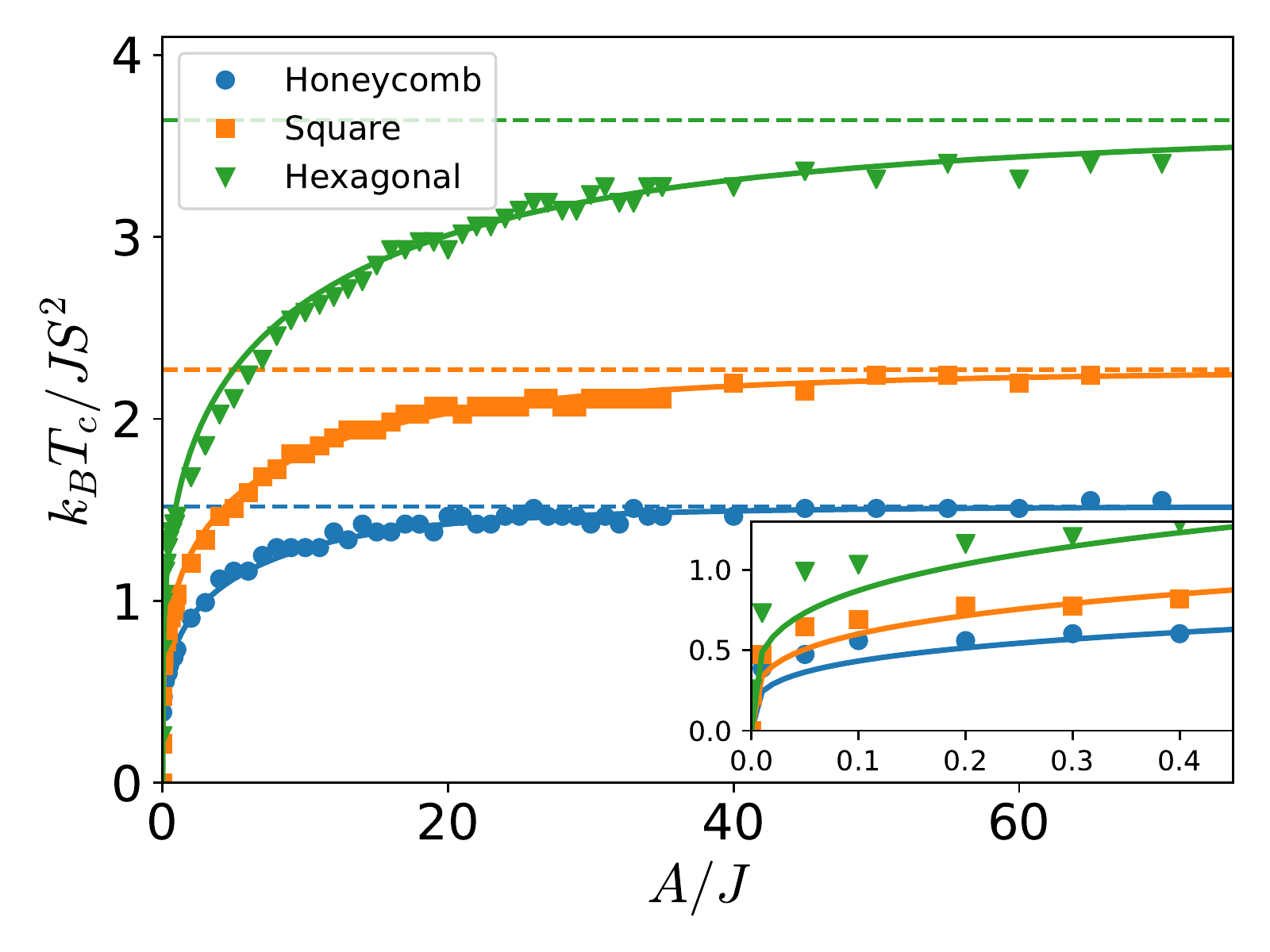}
	\caption{Critical temperature as a function of scaled anisotropy $A/J$ calculated with classical Monte Carlo simulations for the honeycomb, square, and hexagonal lattices with ferromagnetic exchange. The solid lines are obtained from the empirical fitting function Eqs. \eqref{eq:Tc}-\eqref{eq:f}. The Ising limit is indicated by dashed lines for the three lattices.}
	\label{fig:fit}
\end{figure}

We can fit the classical simulations to an analytical function of the form.
\begin{equation}\label{eq:Tc}
T_c = T^\mathrm{Ising}_c f(A/J),
\end{equation}
with
\begin{equation}\label{eq:f}
f(x)=\tanh^{1/4}\Big[\frac{6}{N_{nn}}\log\big(1+\gamma x\big)\Big]
\end{equation}
where $N_{nn}$ is the number of nearest neighbors and $\gamma=0.033$. $T^{\mathrm{Ising}}_c$ is the critical temperatures for the corresponding Ising model, which can be written as $T^{\mathrm{Ising}}_c=S^2J\tilde T_c/k_B$, where $\tilde T_c$ is a dimensionless critical temperature with values of 1.52, 2.27, 2.27, and 3.64 for the honeycomb, quadratic, Kagom\'e, and hexagonal lattices respectively.\cite{Malarz2005} The fit was obtained by noting that the critical temperature has a logarithmic dependence on the anisotropy at low temperature. It is thus natural to base the fit on the $\tanh(x)$ function, which Taylor expands to its arguments for $x\ll1$ and approaches unity for $x\rightarrow\infty$. The exponent of $1/4$, $\gamma$ and the factor of 6 in from of the $\log$ function were obtained by fitting. The comparison of the fit and classical MC simulations for the honeycomb, quadratic, and hexagonal lattices are shown in Fig. \ref{fig:fit}. It should be noted that the fit gives slightly lower values than the simulations at low values of $A/J$. However, for low values of $A/J$ the classical simulations overestimate the critical temperatures compared with RPA and we expect that RPA is more accurate in this limit. Eqs. \eqref{eq:Tc}-\eqref{eq:f} are the main result of the present work and comprises a simple analytical approximation for Curie temperatures of 2D materials with single-ion anisotropy.

\subsection{Anisotropic exchange}
\begin{figure}[b]
	\includegraphics[width=0.45\textwidth]{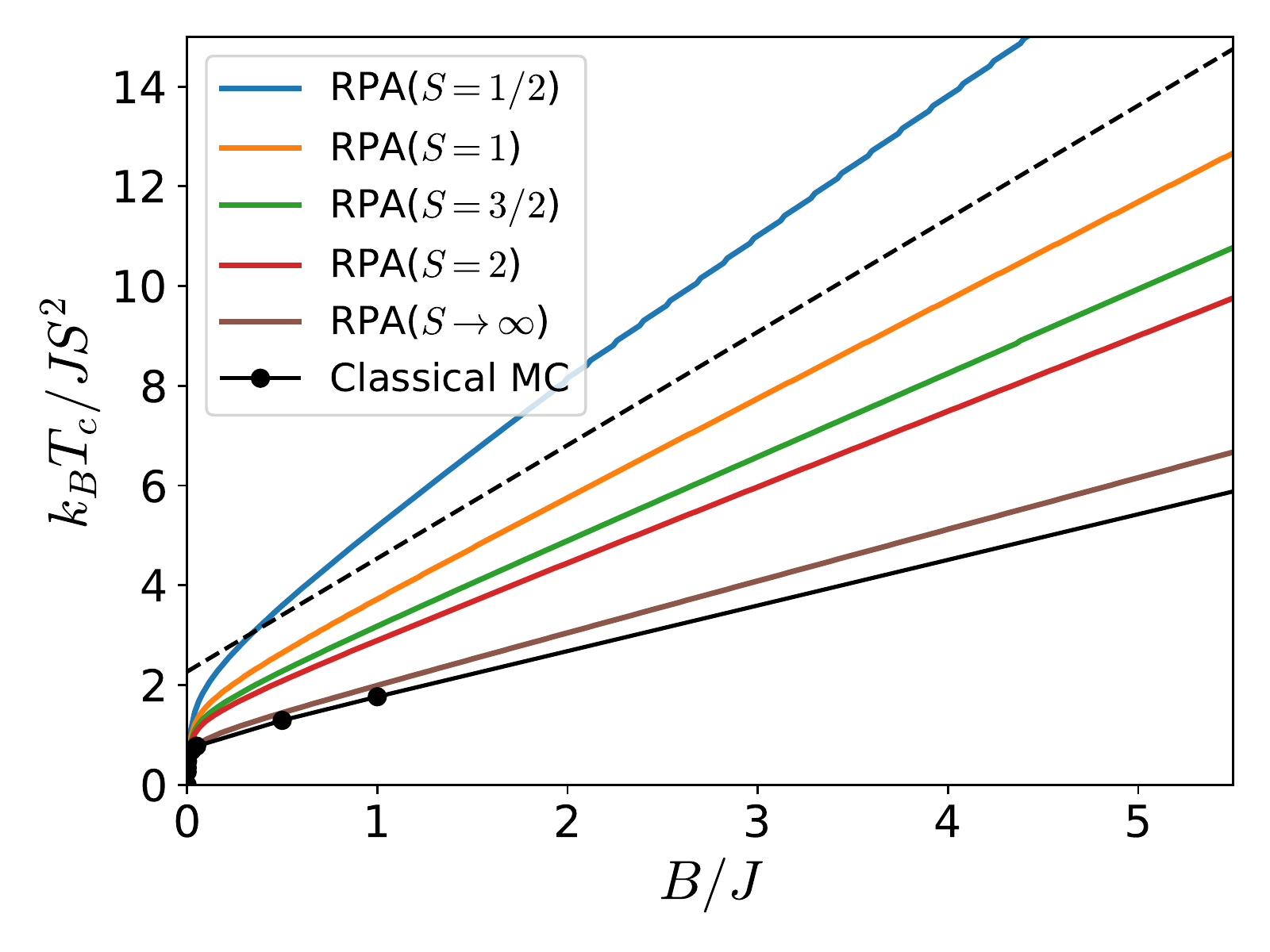}
	\caption{Critical temperature for a quadratic lattice as a function of rescaled anisotropic exchange $B/J$ calculated for $S=1/2$, $S=1$, $S=3/2$, and $S=2$ with ferromagnetic exchange coupling obtained from RPA and MC simulations. The asymptotic Ising limit $T_c=2.27S^2(J+B)$ is indicated by a dashed line.}
	\label{fig:rpa_B}
\end{figure}

The situation become slightly more complicated when anisotropic exchange is also present ($B\neq0$). To exemplify the qualitative differences between anisotropic exchange and single-ion anisotropy we start by considering the case of $A=0$ for a quadratic lattice. The critical temperatures obtained with RPA and MC simulations are shown in Fig. \ref{fig:rpa_B} as a function of the anisotropic exchange parameter $B$. In the case of $B\gg J$ one obtains $H\approx\frac{B+J}{2}\sum S_i^zS_j^z$, which is equivalent to the Ising model when $S=1/2$. However, for $S\geq 1$, the critical temperature is lowered compared to the Ising model due to the non-binary nature of $S^z$. In the limit of $B\gg J$, it may be argued that the values of $\pm S^z$ are favored for all sites, but when the critical temperature is approached the magnetization vanishes in a manner that is quite different from the Ising model, since the magnetization per site is allowed to decrease in addition to the effect of domain formation. From Fig. \ref{fig:rpa_B}, it can be seen that RPA overestimates the asymptotic behavior of $T_c$ as $B/J\rightarrow\infty$ due to the mean-field approximation for correlation effects. However, the classical limit of $S\rightarrow\infty$ is accurately captured by RPA, which agrees well with classical MC simulations. This is in sharp contrast to the case of single-ion anisotropy, where RPA fails completely in the classical limit.

\begin{figure}
	\includegraphics[width=0.45\textwidth]{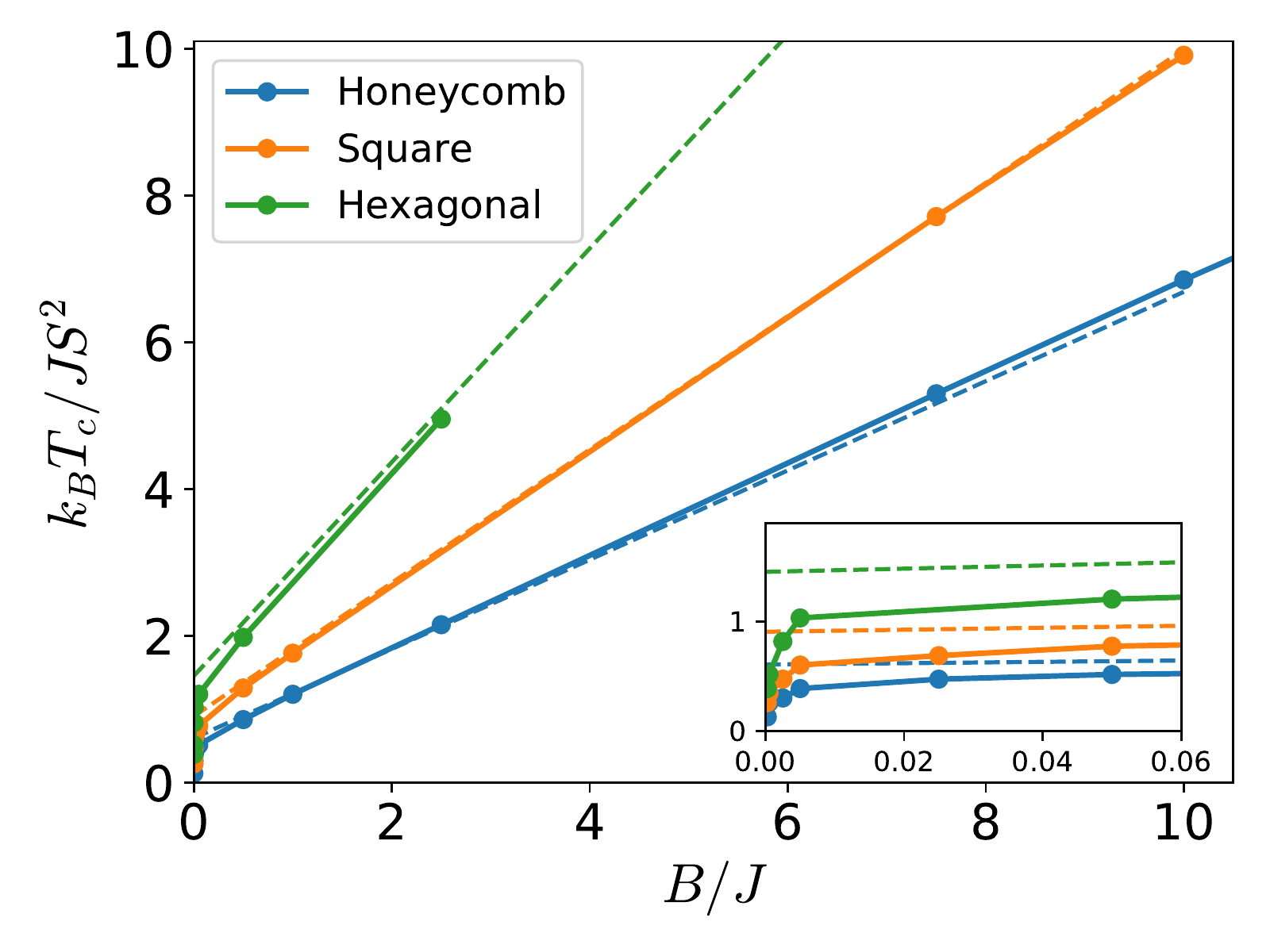}
	\caption{Critical temperature as a function of anisotropy exchange $B/J$ calculated with classical Monte Carlo simulations for the honeycomb, square, and hexagonal lattices with ferromagnetic exchange. The dashed lines indicate the asymptotic Ising limit $T_c^{\mathrm{Ising}}(1+B/J)$ divided by 2.5.}
	\label{fig:classical_B}
\end{figure}

The fact that the model becomes asymptotically equivalent to an Ising type model with continuous spin variables indicate that the asymptotic critical temperatures should resemble those of the Ising model but with an effective value of S that is smaller than the maximum value that enters in the Heisenberg model. For example, taking a spherical average of $S_z$ such that $S^2\rightarrow\langle S_z^2\rangle_\Omega=S^2/3$ one obtains a decrease in in the critical temperature by a factor of three compared to the corresponding Ising model. By inspection of the classical simulations we find that $S^2\rightarrow S^2/2.5$ provides good agreement with our simulations. This is shown in Fig. \ref{fig:classical_B} for the honeycomb, square, and hexagonal lattices. However, we note that this is a strictly classical result and for $S=1/2$ one should reproduce the Ising limit without any rescaling. For finite values of $S$, we expect a rescaling between unity and 2.5.

In order to include the anisotropic exchange in the one-parameter expression \eqref{eq:Tc}, we note that for $B=0$, one can express the single-ion anisotropy in terms of the zero-temperature gap $\Delta$ as $A=\Delta/(2S-1)$ (see Eq. \eqref{eq:eps0}). For $B\neq0$ the situation is more complicated due to the additional dispersion introduced by the anisotropic exchange. Nevertheless, if we choose to regard the zero temperature gap as the primary descriptor for the critical temperature we can write the critical temperature for $S\neq1/2$ as 
\begin{equation}\label{eq:Tc0}
T_c = T^{\mathrm{Ising}}_cf\Big(\frac{\Delta}{J(2S-1)}\Big),
\end{equation}
with
\begin{equation}\label{eq:delta}
\Delta=A(2S-1)+BSN_{nn}
\end{equation}
and $f(x)$ given by Eq. \eqref{eq:f}. 

\subsection{Application to CrI$_3$}
As a test example for the derived expression we consider a monolayer of CrI$_3$, which has recently been shown to exhibit ferromagnetic order below 45 K \cite{Huang2017a}.  We have extracted the Heisenberg model parameters from first principles in the framework of density functional theory. All calculations were performed using the GPAW code \cite{Enkovaara2010a, Larsen2017} and the Perdew-Burke-Ernzerhof (PBE) functional \cite{pbe} for exchange-correlation energies. We have included a Hubbard U correction to properly account for the localization of Cr d-orbitals. A cut-off energy of 800 eV for the plane wave basis and a $\Gamma$-centered Monkhorst-Pack $k$-point with a density of 6 $\mathrm{\AA}^{-1}$ have been used to ensure converged results. Monolayers were separated by 15 {\AA} vacuum and the atomic structure relaxed until all forces declined below 0.01~eV/\AA. 

\begin{figure}
	\includegraphics[width=0.45\textwidth]{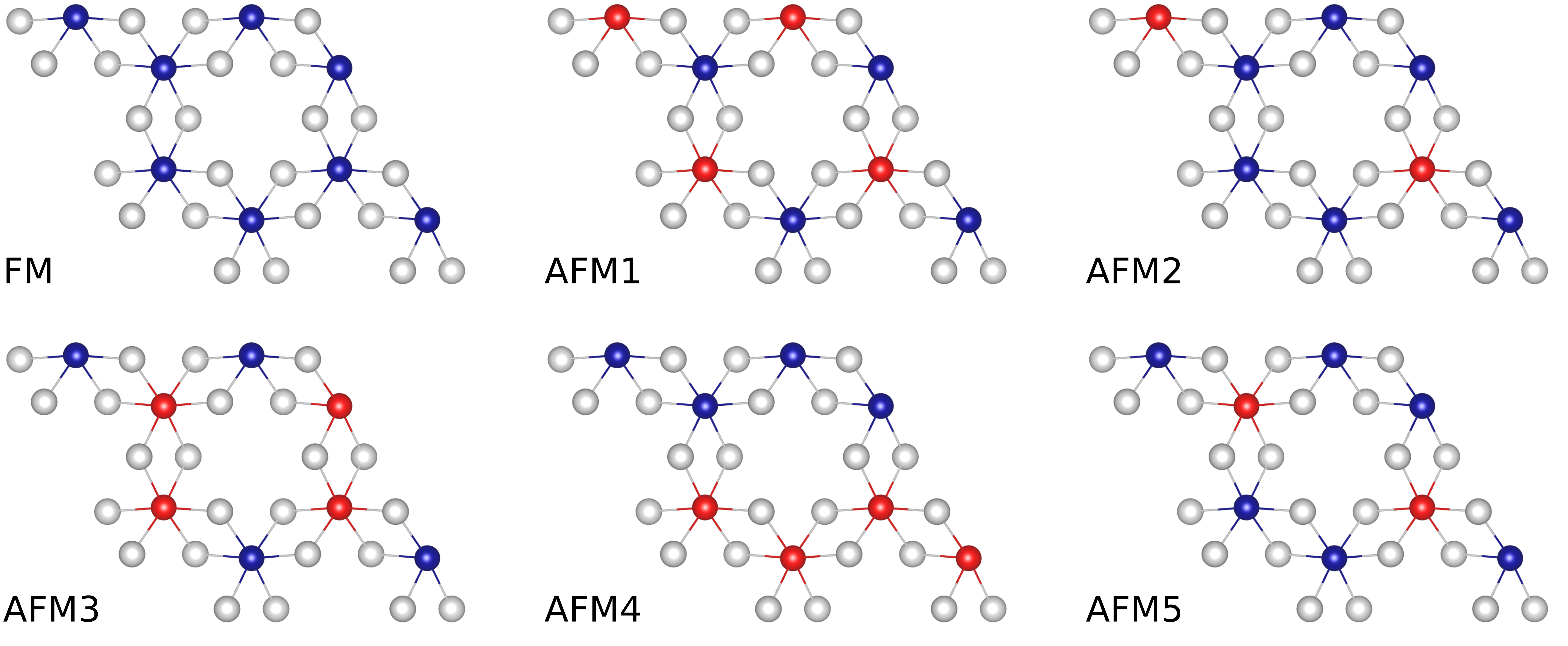}\\
	\includegraphics[width=0.45\textwidth]{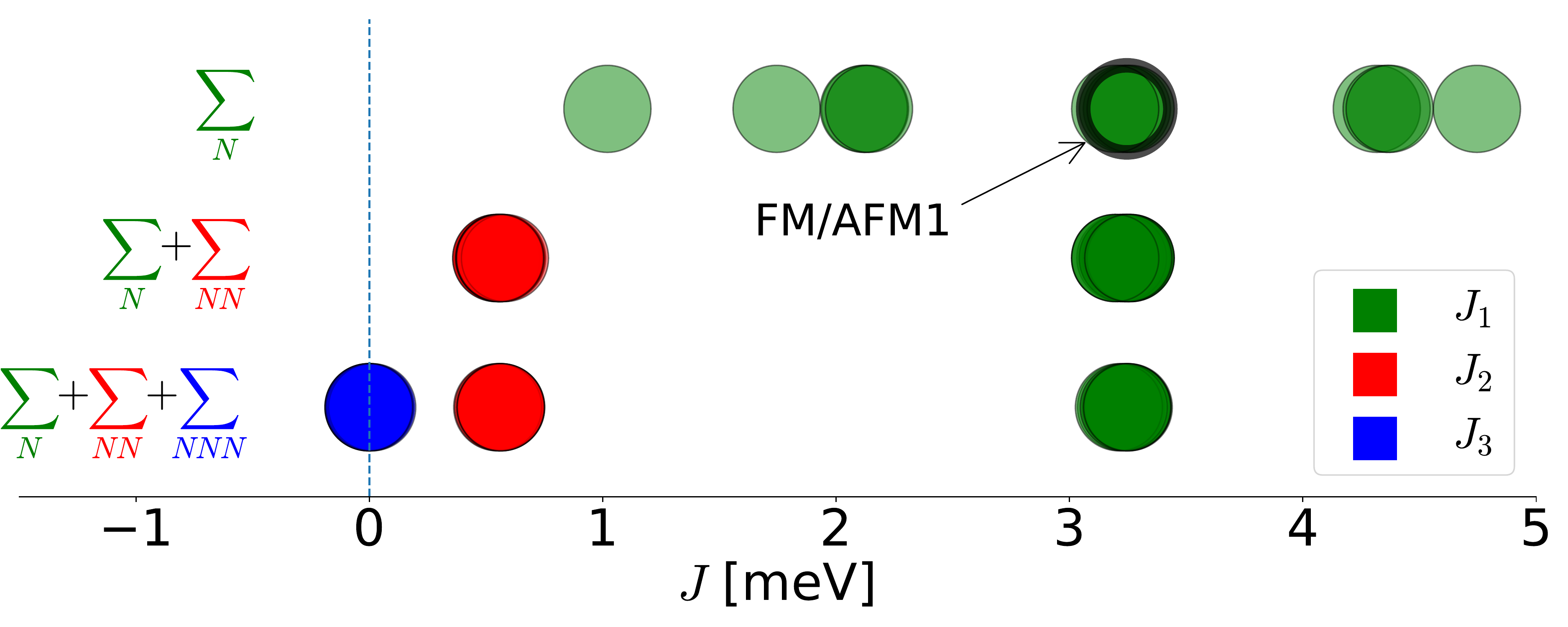}
	\caption{Top: Spin configurations used for the energy mapping analysis with $U=2$ eV. Bottom: Values of $J_1$, $J_2$, and $J_3$ obtained from different combinations of spin configurations. When both second and third nearest neighbor interactions are included the resulting $J_i$ are independent of which spin configurations we use. The black circle shows the result for $J_1$ based on the ferromagnetic and fully antiferromagnetic configurations in a single unit cell}
	\label{fig:conf}
\end{figure}

Magnetic moments of 3 $\mu_\mathrm{B}$ are localized on the Cr atoms and each Cr atom has three nearest neighbors. The exchange coupling constant can be extracted from the energy differences between various spin configurations of the monolayer.\cite{Illas1998, Kodderitzsch2002, Filippetti2005, Xiang2013, Babkevich2016, Olsen2017} In order to calculate the nearest neighbor coupling constant $J_1$, two configurations are required. However, it is important to check that the results are independent of, which configurations are used.  We have thus considered a $2\times2$ unit cell and considered a total of 6 different spin configurations. Taking different combinations of these produce values of $J_1$ that range from 1-5 meV. Including the next nearest neighbor interaction $J_2$ reduces the spread somewhat and give values of $J_1$ in the range 3.1-3.3 meV. Finally including the third nearest neighbor interaction $J_3$ produces converged results that are independent of which spin configurations that we use. Five different combinations of the six spin configurations (four structures are needed to extract three parameters) all produce $J_1=3.24$ meV, $J_2=0.56$ meV and $J_3=0.001$ meV. If we simply consider the two possible magnetic configurations in a single unit cell we obtain $J_1=3.25$ meV (black ring in Fig. \ref{fig:conf}), which is very close to the converged result. Although $J_2$ and $J_3$ are required in order to obtain converged results these parameters are much smaller than $J_1$ and we expect that they will have small influence on the critical temperature. We will thus neglect the second and third nearest neighbor interaction in the following and simply use $J=J_1$ when applying the model \eqref{eq:Tc0}.

The anisotropy parameters can be extracted by a calculations with spin-orbit coupling \cite{Olsen2016a} by comparing the energies in-plane and out-of-plane spin configurations\cite{Lado2017}. In particular
\begin{align}\label{eq:ani}
A=&\frac{\Delta E_\mathrm{FM}+\Delta E_\mathrm{AFM}}{2S^2}\\
B=&\frac{\Delta E_\mathrm{FM}-\Delta E_\mathrm{AFM}}{N_{nn}S^2},
\end{align}
where $\Delta E_\mathrm{FM(AFM)}=E_\mathrm{FM(AFM)}^{(x)}-E_\mathrm{FM(AFM)}^{(z)}$ are the energy differences per atom between in-plane and out-of-plane spin configurations for the ferromagnetic and antiferromagnetic structures and $N_{nn}=3$ is the number of nearest neighbors. Using $U=2$ eV we find
\begin{align}
A^{\mathrm{CrI_3}}=&\;0.056\;\mathrm{meV},\\
B^{\mathrm{CrI_3}}=&\;022\;\mathrm{meV},
\end{align}
and from Eq. \eqref{eq:delta}
\begin{align}
\Delta^{\mathrm{CrI_3}}=&\;1.08\;\mathrm{meV},
\end{align}
which yields
\begin{align}
T_c^{\mathrm{CrI_3}}=&\;42\;\mathrm{K},
\end{align}
from Eqs. \eqref{eq:f}-\eqref{eq:delta}. This is in excellent agreement with the experimental value of 45 K \cite{Huang2017a}. However, the result is somewhat sensitive the the choice of U and calculations with $U=0$, $U=1$, and $U=3$ eV gives $T_c=32$, $T_c=37$, $T_c=47$ K respectively. In general we observe that $J$ and $B$ increases while $A$ decreases, when U is increased and the critical temperature is large linear in $U$ with $dT_c/dU=5$ K/eV. In Ref. \onlinecite{Lado2017} the authors found $A=0$, $B=0.09$ meV, $J=2.2$ meV and $\Delta=0.4$ meV from GGA+U calculations, which is in somewhat disagreement with the present results. In particular, we find that single-ion anisotropy and anisotropic exchange contribute equal amounts to spin gap whereas in Ref. \onlinecite{Lado2017} it was found that single-ion anisotropy is negligible. The reasons for this discrepancy is presently unclear.

\section{Discussion}
To summarize, we have calculated the critical temperatures of various 2D lattices using the RPA and classical MC simulations as a function of single-ion anisotropy and anisotropic exchange. We find that RPA generally tend to overestimate critical temperatures and fails dramatically for large single-ion anisotropies. In contrast, the MC simulations capture the asymptotic Ising limit correctly and agrees reasonably well with RPA at small values of the anisotropy. We used the calculations to obtain an analytical fit for the critical temperature that only depends on the anisotropy constants, nearest neighbor exchange parameter and the (known) critical temperature of the Ising model for the lattices. Since all parameters are easily obtainable from first principles calculations, we expect that the expression will be highly useful for predicting critical temperatures of novel 2D materials as well as verify the microscopic mechanism that underlie magnetic order in experimentally observed 2D ferromagnets.

It should be noted that other approximations to the Heisenberg model, might be better suited than the mean field approximation of the Holstein-Primakoff bosons considered in here. For example, a mean field treatment of the Schwinger bosons approach\cite{Auerbach} has been successfully applied to describe the properties of ferrimagnetic systems\cite{Wu} and antiferromagnetic Kagome lattices\cite{Mondal} and it would be very interesting to investigate the performance of such methods in future works.

The present results obtained from classical Monte Carlo simulations are largely based on classical simulations and are not expected to be valid for the the important case of $S=1/2$ systems. Even for $S=1$ and $S=3/2$, it is hard to argue that a classical approach is reliable and the only evidence for the validity so far is the good agreement with the experimentally observed Curie temperature for CrI$_3$. To verify the reliability of the method it would thus be highly desirable to obtain accurate results based on either many-body techniques beyond the RPA \cite{Pershoguba2018} or quantum Monte Carlo simulations of the anisotropic Heisenberg model. Moreover, the first principles evaluation of the Heisenberg parameters seems to be rather sensitive to the exact approach used to calculate them and the exact values of the parameters are still debated.\cite{Lado2017} Whereas spin waves\cite{Samuelsen1971} and critical exponents\cite{Liu2018} have been measured in bulk CrI$_3$ there is not yet any direct measurements for a monolayer of CrI$_3$ and it is currently not possible to determine the Heisenberg parameters experimentally. 

Presently, there is very few known purely 2D magnetic materials, which severely limits the possibilities of benchmarking ab initio calculations and models for magnetism in 2D against experimental observations. However, with the rapid developments in synthesis, characterization, and prediction of 2D materials,\cite{Zhou2018, Mounet2018, Haastrup2018} it is likely that several new magnetic 2D materials will be observed in the near future.


%

\end{document}